\documentstyle[aps,psfig]{revtex}
\begin{document}
\twocolumn[\hsize\textwidth\columnwidth\hsize\csname
@twocolumnfalse\endcsname
\title{Robustness as an Evolutionary Principle$^*$}
\author{Stefan Bornholdt$^a$ and Kim Sneppen$^b$}
\address{$^a$Institut f\"ur Theoretische Physik,
Universit\"at Kiel, Leibnizstrasse 15, D-24098 Kiel, Germany
\\
$^b$ NORDITA, Blegdamsvej 17, DK-2100 Copenhagen, Denmark 
\\
$^*$Submitted to Proc.\ R.\ Soc.\ London B
}
\maketitle
\begin{abstract}
We suggest to simulate evolution of complex organisms 
constrained by the sole requirement of robustness in 
their expression patterns. This scenario is illustrated by 
evolving discrete logical networks with epigenetic properties. 
Evidence for dynamical features in the evolved networks
is found that can be related to biological observables.
\end{abstract}
\bigskip
]

\section{Introduction} 
A common concept in evolution is fitness and fitness landscapes 
\cite{Wright},
and often evolution is viewed as hill climbing, possibly
with jumps between fitness maxima \cite{Newman,Lande}.
However, fitness landscapes implicitly assume that fitness
is varying over a well-defined metric in genomic space.
This would be the case if single point mutations were a driving force.
However, significant genome rearrangements are observed 
already in the rather brief real-time evolution experiments
of {\sl Escherichia coli} cultures of Papadopoulos et al. \cite{Lanski}.
Genomic rearrangements short-circuit the simple metric
generated by one point mutations, usually underlying the 
intuition of evolution on landscapes. 
As a consequence the combinatorial distance for moving 
from a genome A to a genome B
may easily be different from the distance of the opposite move, 
simplest exemplified by deletions and insertions.
Thus, although fitness landscapes 
have a meaning for the small scale adjustments associated to
fine-tuning of binding constants, it is an unjustified concept for
evolutionary changes on the scale of speciation events.
 
Abandoning fitness landscapes we here instead discuss 
the possibility that evolution progresses through a process 
where genotypes and phenotypes 
subsequently set the frame at which the other may change.
Of particular relevance for this view of evolution 
is the fact that one often observes different phenotypes 
for the same genotype. 
This viewpoint is in part supported by cell 
differentiations within one organism, in part 
supported with epigenetics and the
large class of organisms which 
undergo metamorphosis and thus exist in several 
phenotypes for the same genotype. 
Recently, it has also been proposed that
genotype-phenotype ambiguity \cite{Kaneko} 
is governing speciation events.

A class of systems that exhibits epigenetics is represented by 
the logical networks,
where nodes in the network take values on or off,
as function of the output of specified other nodes.
This has been suggested to model the regulatory
gene circuits \cite{Kauffman,Somogyi,Thieffry98} 
where specific genes may or may not be 
expressed as function of other genes.
In terms of these models it is natural to define genotypes in form of
the topology and rules of the nodes in the network.
The phenotypes are similarly associated to the dynamical expression
patterns of the network. 

To define the rules under which phenotypes and genotypes
set the frame for each other's development, a model for 
evolution should fulfill the requirement of robustness.
Robustness is defined as the ability to function in face of
substantial change in components \cite{Savageau,Hartwell,Barkai,Little}. 
Robustness is an important ingredient in simple molecular networks 
and probably also an important feature of gene regulation on both,  
small and large scale. 
In terms of logical networks, robustness is implemented by constraining
subsequent networks to have similar expression patterns. 

This article is organized as follows:
First we discuss dynamics on logical networks and 
numerically review the basic properties of attractors 
of random threshold networks and Boolean networks.
Then we propose a minimal evolution model and 
investigate its statistical and structural implications
for the evolved networks. 
Finally biological implications, and possible experimental 
approaches to the dynamics of real genetic networks are discussed.

\section{Dynamics on logical networks} 
Let us first discuss two prototype networks that exhibit epigenetics, 
Boolean networks \cite{Kauffman,Somogyi} and threshold networks 
\cite{threshold}. These are both networks of logical functions 
and share similar dynamical properties. We here briefly describe their 
definition and dynamical features.
In both networks each node is taking one of two discrete 
values, $\pm 1$, that at each timestep is a discrete function 
of the value of some fixed set of other nodes specified by a 
wiring diagram.
If we denote the links that provide input to node $i$ by $\{w_{ij}\}$, 
with $w_{ij}=\pm1$ also, 
then for the threshold network case the updating rule is additive
\begin{eqnarray}
\sigma_i \; = 1 \;\;\; if \;\;\; 
\sum_{j\in \{w_i\}} w_{ij} \sigma_j \; \ge 0 \\
\sigma_i \; = -1 \;\;\; if \;\;\; 
\sum_{j\in \{w_i\}} w_{ij} \sigma_j \; < \; 0 
\end{eqnarray}
In the Boolean network case the updating is a general
Boolean function of the input variable
\begin{equation}
\sigma_i = B( {\sigma_j}'s \; which\; provide \; input \; to\; i ).  
\end{equation}
Thus, the threshold networks form a hugely restricted set of the 
Boolean networks. Boolean networks include all 
nonlinear combinations of input nodes, including functions 
as, for example, the ``exclusive or''. 

The basic property of logical networks
is a dynamics of the state vector \{ $\sigma_i$ \}
characterized by transients that lead to subsequent attractors.
The attractor length depends on the topology of the network.
Below a critical connectivity $K_c\sim 2$ \cite{Derrida,Kauffman} the network
decouples into many disconnected regions, 
i.e., the corresponding genome expression would become modular,
with essentially independent gene activity.
Above $K_c$ any local damage will initiate an avalanche of activity
that may propagate throughout most of the system.
For any $K$ above $K_c$ the attractor period diverges exponentially
with respect to system size $N$ and in some
interval above $K_c$ the period length in fact also increases 
nearly exponentially with connectivity $K$ \cite{Bastola}.

\section{Structural evolution of networks}  
Dynamics may occur on networks as defined by the rule above,
but at least as important is the dynamics of network topology.
In terms of network topology an evolution means
a change in the wiring
$\{w_{ij} \} \rightarrow \{w_{ij}' \}$ that takes
place on a much slower timescale than the $\{\sigma_j \}$ updating.
The evolution of such networks represents the extended degree
of genetic network engineering that seems to be needed to account
for the large differences in the structure of species genomes \cite{Shapiro},
given the slow and steady speed of single protein evolution \cite{Kimura}.

We have in an earlier publication proposed to evolve 
Boolean networks with the sole constraint
of continuity in expression pattern \cite{prl}.
Here we simplify this model by simple damage spreading testing: 
\vspace{0.5cm}

\noindent
{\bf The model evolves a new single network
from an old network by accepting rewiring mutations 
with a rate determined by expression overlap}.
\vspace{0.5cm}

This is a minimal constraint scenario with no outside fitness 
imposed. 
Further the model tends to select for networks
which have high overlap with neighbor mutant networks,
thus securing robustness.

Now let us formulate an operational version of the
evolution in terms of threshold networks as these
have comparable structural and statistical
features to the Boolean ones \cite{threshold}.
Consider a threshold network with $N$ nodes. 
To each of these let us assign a logical
variable $\sigma_i=$ $-1$ or $+1$.
The states $\{ \sigma_i \}$ of the $N$ nodes are simultaneously
updated according to (1) where the links $w_{ij}$ are
specified by a matrix.
The entry value of the connectivity matrix
$w_{ij}$ may take values $-1$ and $+1$ in case of a link between
$i$ and $j$, and the value $0$ 
if $i$ is not connected to $j$. 

The system that is evolved is the set of couplings 
$w_{ij}$ in a single network. One evolutionary time step of the network is:
\vspace{0.2cm}

\noindent
{\bf 1)} Create a daughter network by 
a) adding, b) removing, or c)
adding and removing a weight in the coupling matrix $w_{ij}$ at random,
each option occurring with probability $p=1/3$.
This means turning a $w_{ij}=0$ to a randomly chosen $\pm 1$ or vice versa.

\noindent
{\bf 2)} Select a random input state $\{ \sigma_i \}$.
Iterate simultaneously both the mother and the daughter system 
from this state until they 
either have reached and completed the same attractor cycle,
or until a time where $\{ \sigma_i \}$ differs between the two networks.
In case their dynamics is identical then replace the mother with the daughter network.
In case their dynamics differs, keep the mother network.
\vspace{0.2cm}

Thus, the dynamics looks for mutations which
are phenotypically silent, i.e., these are
neutrally inherited under at least some external condition.
Notice that adding a link involves selecting a new $w_{ij}$,
thus changing the rule on the same timescale as the network connectivity.
Iterating these steps represents an evolution which
proceeds by checking overlap in expression pattern between networks.
If there are many states $\{ \sigma_i \}$
that give the same expression of the two networks,
then transitions between them are fast.
On the other hand, if there are only very few states $\{ \sigma_i \}$
which result in the same expression for the two networks,
then the transition rate from one network to the other is small. 
If this is true for all its neighbors then the evolutionary process will
be hugely slowed down. 

In Fig.\ 1 the connectivity
change with time for a threshold network of size $N=32$ is shown.
\begin{figure}[htb]
\let\picnaturalsize=N
\def\picsize{85mm}
\def\picfilename{fig1.ps}
\ifx\nopictures Y\else{\ifx\epsfloaded Y\else\input epsf \fi
\let\epsfloaded=Y
\centerline{\ifx\picnaturalsize N\epsfxsize \picsize\fi
\epsfbox{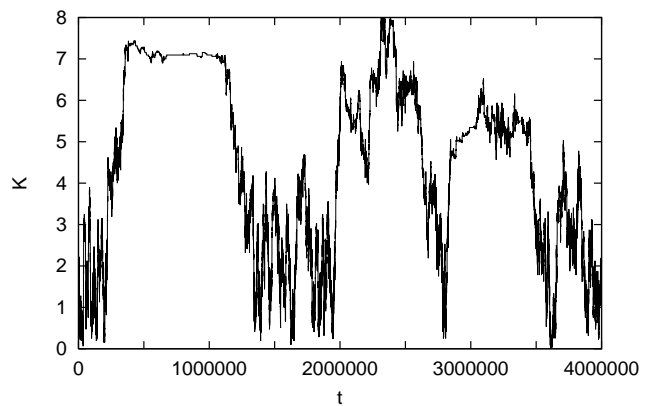}}}\fi
\caption{
Long time evolution for the connectivity of
of a threshold network with N=32 nodes. Connectivities are constrained 
to be below $K=8$. One observes long periods of stasis 
interrupted by sudden changes, 
reminiscent of punctuated equilibrium.
}  
\end{figure}
Time is counted as number of attempted mutations, 
and one observes that especially for high connectivity 
the system may stay long time
at a particular network before an allowed mutation leads
to punctuations of the stasis.
The overall distribution of waiting times is $\sim 1/t^{2\pm0.2}$.

One feature of the evolution is the
structure of the evolved networks, which can be 
quantified by the average length of attractors for 
the generated networks.
This is shown in Fig.\ 2, where they are compared  
with attractor lengths for random networks at the same connectivity.
\begin{figure}[htb]
\let\picnaturalsize=N
\def\picsize{85mm}
\def\picfilename{fig2.ps}
\ifx\nopictures Y\else{\ifx\epsfloaded Y\else\input epsf \fi
\let\epsfloaded=Y
\centerline{\ifx\picnaturalsize N\epsfxsize \picsize\fi
\epsfbox{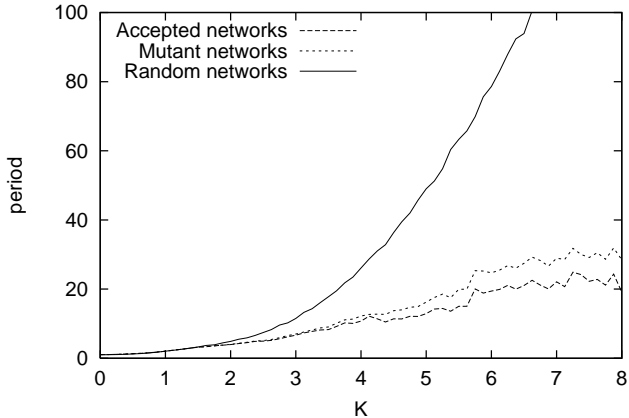}}}\fi
\caption{
Average length of periodic attractors for evolved and for random networks. 
Also the periods of the unsuccessful mutations 
in the presence of newly chosen random initial conditions 
are shown, demonstrating 
that selection of networks is indeed operating in structure space and 
the specific input configuration in the event of selection does not play 
a major role.}  
\end{figure}
One observes that the evolved networks have much shorter attractors
than the random ones, thus our evolution scenario
favors simplicity of expression. 

To examine further the expression behavior of the networks
let us consider the size of frozen components as introduced by Kauffman
for Boolean networks \cite{Kauffman}.
A frozen component is the set of nodes connected to a given attractor
that does not change at any time when you iterate along the attractor,
i.e., a frozen component represents 
genes which are anesthesized under a given attractor/initial conditions.
In Fig.\ 3 one sees that the frozen component for the evolved network 
typically involves half the system, and thus is much larger than 
the typical frozen component associated to attractors 
of randomly generated threshold networks.
Also we test frozen components for random one mutant neighbors 
of the selected ones, and find that these networks also have huge
frozen components. 

Let us finally look at the active part of the network
and the complexity of its expression pattern. As a quite large fraction
of the nodes may belong to the frozen component of the network, 
the remaining active part of the nodes may behave differently  
from the average dynamics of the whole network.  
One possible measure is the number of times, each non-frozen node 
switches its state during the dynamical attractor.  
\begin{figure}[b]
\let\picnaturalsize=N
\def\picsize{85mm}
\def\picfilename{fig3.ps}
\ifx\nopictures Y\else{\ifx\epsfloaded Y\else\input epsf \fi
\let\epsfloaded=Y
\centerline{\ifx\picnaturalsize N\epsfxsize \picsize\fi
\epsfbox{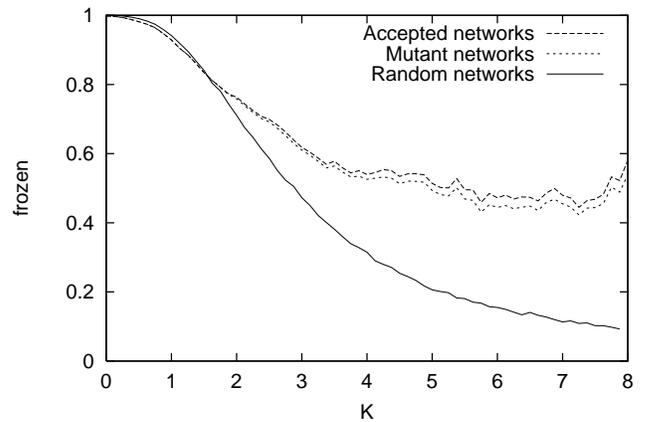}}}\fi
\caption{
Average size of frozen components as a function of connectivity for 
evolved and random networks. 
The frozen component is the set of all 
nodes that do not switch during the attractor.   
One observes that the robustness constraint in evolution 
favors a larger frozen component.}  
\end{figure}
In Fig.\ 4 this quantity is shown for random networks as well as 
evolved networks. One observes that the active part of the 
evolved networks exhibits a much simpler expression pattern 
than that of a random network of comparable connectivity.  
\begin{figure}[htb]
\let\picnaturalsize=N
\def\picsize{85mm}
\def\picfilename{fig4.ps}
\ifx\nopictures Y\else{\ifx\epsfloaded Y\else\input epsf \fi
\let\epsfloaded=Y
\centerline{\ifx\picnaturalsize N\epsfxsize \picsize\fi
\epsfbox{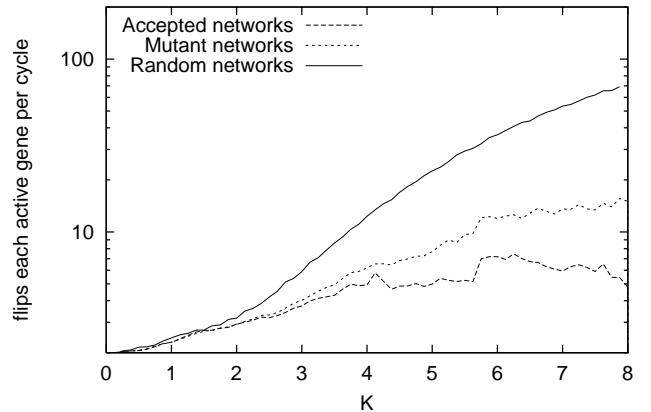}}}\fi
\caption{
Average number of flips per node in the non-frozen part of the network,  
as a function of connectivity for evolved and for random networks. 
The evolved networks show a reduced activity in the non-frozen 
nodes resulting in simple expression patterns as compared to those of 
random networks of same connectivities. Notice that the number 
counts off-on and on-off transitions of the nodes as separate events. 
}  
\end{figure}

Overall, requiring robustness as an evolution criterion 
has observable consequences for both, 
the temporal evolution pattern, 
and for confining possible genetic network architectures
to the ones with simple expression patterns.

\section{Discussion} 
Some quantitative testing of the minimal evolution scenario
is possible on the macro-evolutionary scale.
Here the intermittent evolution of the networks bears resemblance to the 
punctuated equilibrium observed for species in the fossil record \cite{GE1993}.
Quantitatively the $1/f$ power spectra and $1/t^2$ stability distribution
for single networks, that one finds for this model as well as for
the earlier version \cite{prl},
compares well with the similar scalings observed for the
statistics of birth and death of individual species in the evolutionary record 
\cite{Bak-Sneppen,Sole}. Obviously the here ignored features related to co-evolution 
prevent us from discussing co-extinctions \cite{Bak-Sneppen}. 
In fact the analogy can even be fine-grained into a sum of 
characteristic lifetimes,
each associated to a given structural feature of the networks \cite{prl}.
A similar decomposition is known from the fossil record \cite{VanValen}, 
where groups of related species display Poisson
distributed lifetimes and therefore similar 
evolutionary stability.

A validation on the microlevel based on statistical properties of
genetic regulatory circuits has to be based either on 
properties of genetic networks \cite{Somogyi}
or on evolution and mutation
experiments of fast lived organisms as {\sl E.\ coli} \cite{Lanski}.
A key number is the estimated average connectivity
$K$ of $2\rightarrow 3$ in the {\sl E.\ coli} genome \cite{Thieffry}. 
Information on the overall organization of these 
genetic networks is obtained from gene knock out experiments.

A quantitative support for a connected genome
can be deduced from Elena and Lenski's \cite{Elena} experiments on
double mutants, which demonstrated that about 30-60\%
of these (dependent on interpretation)
change their fitness in a cooperative manner.
In terms of our networks, we accordingly should expect a 
coupled genetic expression for about half of the
of pairs of genes. Although our evolved networks
can give such correlations for
the connectivity estimate of 2-3 given by \cite{Thieffry},
the uncertainty is still so large that
random networks also are in accordance with data.
Further one should keep in mind that
the {\sl E.\ coli} genome is large and not well represented
by threshold dynamics of all nodes, 
and also that only between 45 and 178
of the {\sl E.\ coli}'s 4290
genes are likely to mediate regulatory functions \cite{Ecoli}. 
Thus, most of the detected gene-gene correlations 
presumably involve genes which are not even regulatory,
but instead metabolic and their effect on each other 
more indirect than in the case of the regulatory ones.
Presumably one would obtain stronger elements of both coupling and 
correlation if one specialized on regulatory genes.
Thus one may wish for experiments where
one and two point mutations are performed in regulatory genes only.
A more direct test of our hypothesis of damage control 
as a selection criterion may be obtained from 
careful analysis of the evolution of gene regulation in 
evolving {\sl E.\ coli} cultures.

Another interesting observation is the simplicity 
of biological expression patterns. 
For example as observed in yeast many genes 
are only active one or two times during the 
expression cycle \cite{yeast}, 
thus switching from off to on or on to off 
occurs for each gene in this system only a 
few times during expression. 
For random dynamical networks of comparable 
size one would expect a much higher activity.  
Thus surprisingly simple expression patterns are 
observed in biological gene regulatory circuits. 
This compares well with 
our model observation where 
simplicity of expression patterns emerges 
as a result of the evolutionary constraint.

\section{Summary}
In this article we have proposed a computer simulation
of evolution operating on logical networks. 
The scenario mimics an evolution 
of gene regulatory circuits that is 
governed by the requirement of robustness only.
The resulting dynamics evolves networks which have
very large frozen components and short attractors. 
Thus they evolve to an ordered structure that 
counteracts the increasing chaos
when networks become densely connected.
The evolved architecture is characterized by 
simplicity of expression pattern and 
increased robustness to permanent mutational fluctuations 
in the network architecture -- features
that are also seen in real molecular networks.
\vspace{1.0cm}

{\bf Acknowledgment} We thank Stanley Brown 
for valuable comments on the manuscript.

\end{document}